\documentclass[aps,prl,twocolumn,floats]{revtex4-1}
\usepackage{graphicx}
\usepackage{color}
\usepackage{bm}
\usepackage{csquotes}
\usepackage{footmisc}
\makeatletter

\newcommand{\Rmnum}[1]{\expandafter\@slowromancap\romannumeral #1@}
\makeatother

\begin{document}

\title{\bf Designing multi-level resistance states for multi-bit storage using half doped manganites}

\author{Sanjib Banik}
\affiliation{CMP Division, Saha Institute of Nuclear Physics, Kolkata 700064, India}
\author{Kalipada Das}
\affiliation{CMP Division, Saha Institute of Nuclear Physics, Kolkata 700064, India}
\author{Kalpataru Pradhan}
\email{kalpataru.pradhan@saha.ac.in}
\affiliation{CMP Division, Saha Institute of Nuclear Physics, Kolkata 700064, India}
\author{I. Das}
\email{indranil.das@saha.ac.in}
\affiliation{CMP Division, Saha Institute of Nuclear Physics, Kolkata 700064, India}

\begin{abstract}
Designing nonvolatile multi-level resistive devices is the necessity of time to go
beyond traditional one-bit storage systems, thus enhancing the storage density.
Here, we explore the electronic phase competition scenario to design multi-level
resistance states using a half doped CE-type charge ordered insulating bulk manganite,
${Sm_{0.5}Ca_{0.25}Sr_{0.25}MnO_3}$ (SCSMO). By introducing electronic phase coexistence
in a controllable manner in SCSMO, we show that the system can be stabilized into
several metastable states, against thermal cycling, up to 62 K. As a result the
magnetization (and the resistivity) remains unaltered during the thermal cycling.
Monte Carlo calculations using two-band double  exchange model, including super-exchange,
electron-phonon coupling and quenched disorder, show that the system freezes into a
phase coexistence metastable state during the thermal cycling due to the chemical disorder
in SCSMO. Using the obtained insights we outline a pathway by utilizing four reversible
metastable resistance states to design a prototype multi-bit memory device.
\end{abstract}


\maketitle


Fabrication of non-volatile memory (NVM) devices with high storage density, fast
switching speed and low power consumption is of current research
interest~\cite{Scott,Meijer,Chen,Pan,Hwang}. Present day NVM devices mostly use random
access memories (RAM), based on resistive switching (RS) phenomenon~\cite{Sawa,Jeong,Kim,
Ielmini,Russo}. A typical resistive memory device uses two resistive states to store
binary digits 0 and 1. Such memory devices based on resistive switching between two states
face scaling issues to deal with next generation computing systems~\cite{Wang1}.

A promising and challenging direction to overcome the scaling issues is to design
multi-bit memory devices where the intermediate resistivity states can also be used to
store digital data. For example a four-level resistive system can store two-bit
data (00 or 01 or 10 or 11) in each chips as compared to one-bit data (0 or 1) in
prototype two-level resistive system. The feasibility of multi-level cell has been
investigated for phase change memory (PCM) in chalogenides~\cite{Wuttig,Koelmans,Wang,Salinga,Xie}.
In PCM cells the crystallinity is controlled to obtain different resistance states.
In spite of having tremendous functionality, the elemental segregation upon repeated
cycling remains a major drawback for PCM cells~\cite{Raoux1,Raoux2,Beneventi,Simpson,Boniardi,Chen1}.
In this context, the obvious question arises: can one design multiple resistance states
by controlling the electronic phases, commonly seen in phase coexisted transition metal
oxides~\cite{Asamitsu,Uehara,Ahn,Sarma,Odagawa,Rubi,Hoffman} to overcome
the problem of elemental segregation?

In the present communication, we report a multi-level resistive system by tuning the phase
competition in a half doped manganites. The charge-ordered antiferromagnetic insulating
(CO-AFM-I) state, in
our ${Sm_{0.5}Ca_{0.25}Sr_{0.25}MnO_3}$ (SCSMO) sample, melts to a ferromagnetic (FM)
metallic state at a moderate magnetic field 70 kOe (one can use any field greater than
critical field value $\sim$45 kOe~\cite{Sanjibnpg}) at 10 K and the system converts to a
phase coexisted metallic state after the removal of the magnetic field. During the zero
field warming, we stop at four representative
temperatures (35 K, 45 K, 52 K, 58 K) and perform temperature cycling between each
representative temperatures and 10 K.
Distinctly different magnetization and resistivity data generated at above four representative
temperatures remains more or
less unchanged during the temperature cycling. Our Monte Carlo
calculations using two-band double-exchange model verifies that the system freezes to a
metastable phase during the thermal cycling and attributes it to the chemical disorder.
In addition, we show that these metastable states are fully reproducible to use in
memory devices. Then, using four such representative metastable states (in principle
many more states can be generated) we outline a procedure to design a prototype multi-bit
storage device.


\subsection*{\noindent Magnetic and magnetotransport measurements.}
\noindent

\begin{figure*}
\includegraphics[width=0.95\textwidth]{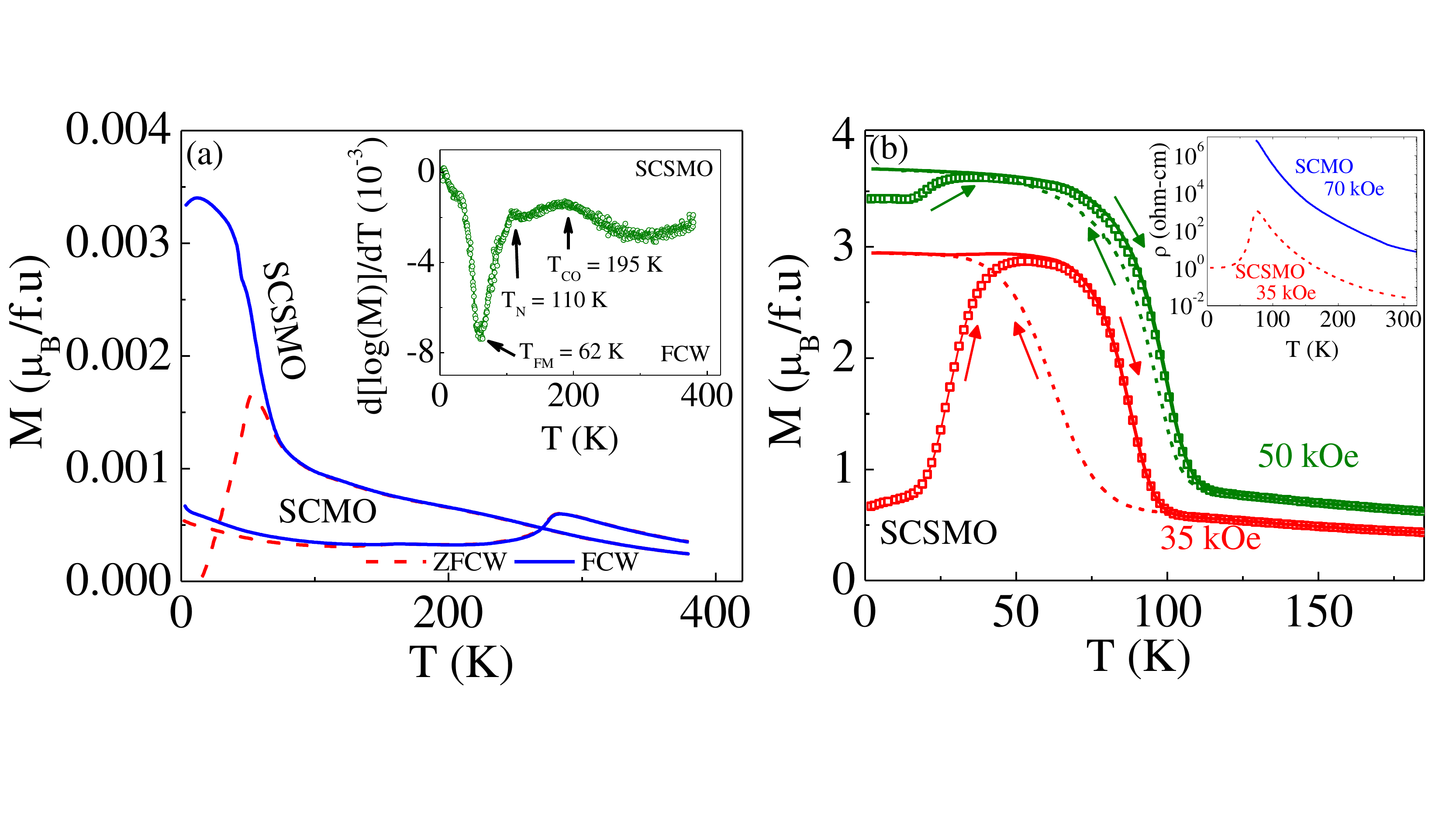}
\centering \caption[ ]
{Magnetic properties of SCSMO: (a) Temperature dependence of magnetization measured in
the presence of 100 Oe magnetic fields using ZFCW (dotted line) and FCW (dashed line)
protocol. Different transition temperatures are marked by the arrows ($T_{CO}$ = 195 K,
$T_N$ = 110 K and $T_{FM}$ = 62 K) for SCSMO in the inset. Magnetization vs.
temperature is also plotted for SCMO. (b) Magnetization with temperature for 35 kOe
and 50 kOe magnetic fields (dotted line, solid line and line
with square symbols indicate the FCC, FCW and ZFCW magnetization data, respectively).
Inset in (b) shows resistivity vs. temperature for SCMO (at 70 kOe) and SCSMO (at 35 kOe)
}
\label{m-temp}
\end{figure*}

We prepared high quality polycrystalline $Sm_{0.5}Ca_{0.25}Sr_{0.25}MnO_3$ (SCSMO)
sample using well-known sol-gel technique (please see the method section for details).
Fig.~\ref{m-temp}(a) shows the magnetization ($M$) vs. temperature ($T$), measured in
the presence of 100 Oe external magnetic field for SCSMO [and $Sm_{0.5}Ca_{0.5}MnO_3$ (SCMO)]
using zero field cooled warming
(ZFCW), field cooled cooling (FCC), and field cooled warming (FCW) protocols. FCW curve in
both SCSMO and SCMO exactly overlay the FCC curve. The bifurcation between
ZFCW and FCW curves in SCSMO as well as peak in ZFCW indicates that small but finite
ferromagnetic fractions are induced in SCSMO which is absent in SCMO. This is
due to the substitution of larger Sr ions in place of Ca in the parent compound
${Sm_{0.5}Ca_{0.5}MnO_3}$~\cite{Sanjibnpg}. The dip in the  $dM/dT$ vs. $T$ curve for SCSMO
[see the inset of Fig.~\ref{m-temp}(a)] also substantiate the presence of ferromagnetic
fractions below 62 K.
Along with the ferromagnetic transition the inset also depicts an antiferromagnetic
transition $(T_N)$ at 110 K and a charge ordered transition $(T_{CO})$ around 195 K.
All these results signifies the presence of phase coexistence
at low temperatures.

To further study the phase coexistence in SCSMO, we apply 35 kOe magnetic field
and measure $M$ vs. $T$
using three (ZFCW, FCW and FCC) protocols, as shown in Fig.~\ref{m-temp}(b). Here ZFCW
curves also shows a maxima around 60 K similar to ZFCW curve in 100 Oe magnetic field.
For SCSMO, 35 kOe magnetic field melts the CO-AFM-I state to FM metallic state
although 70 kOe magnetic field had no effect on the parent compound SCMO (see the inset
for resistivity comparison).
The moderate hysteresis between the FCC and the FCW measurements in 35 kOe indicates the
phase coexistence between the FM and The AFM phases over the temperature range 50--100 K, but
the phase coexistence diminishes for 50 kOe magnetic field [see Fig.~\ref{m-temp}(b)].

\begin{figure*}
\includegraphics[width=0.85\textwidth]{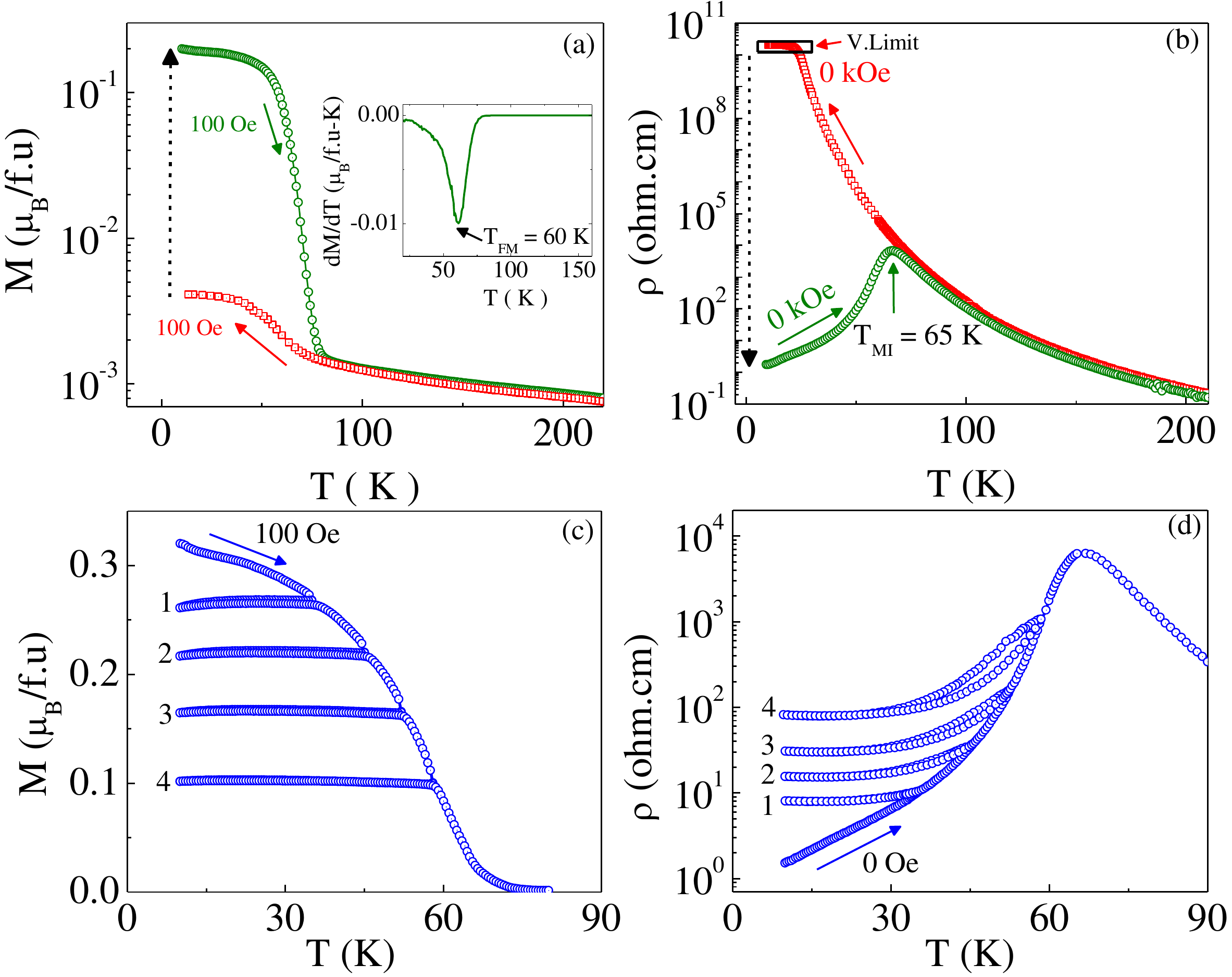}
\centering \caption[ ]
{\label{m-steps} Multi-level resistive states in SCSMO:
(a) Square symbols (FCC): temperature dependence of the magnetization in 100 Oe magnetic
field. Circle symbols (ZFW-I): variation of magnetization with temperature in 100 Oe
magnetic field (after applying and removing 70 kOe magnetic field at 10 K). Inset shows
the temperature derivative of $M(T)$ data taken for the warming process.
(c) Similar to ZFW-I, but temperature cycling is performed while increasing the
temperature at four representative temperature points (defined as ZFW-II).
Resistivity corresponding to magnetization data in (a) and (c) are plotted (b) and (d)
respectively. V-limit is limiting value of our measuring instruments (see
method section for details).}
\end{figure*}

In order to explore the evolution of phase coexistence with the temperature, we measure
$M$ vs. $T$ and $\rho$ vs. $T$ using different protocols. Initially we cooled the system
from 300 K to 10 K. For magnetization measurements a test field (100 Oe) is always applied
unless otherwise specified. Magnetization at 10 K remains very small ($<10^{-2}$ $\mu_B$/f.u)
and resistivity is found to be insulting in nature [see Fig.~\ref{m-steps}(a)and (b))]. At
this point, we apply and then remove the 70 kOe magnetic field, and record the magnetization
(the resistivity) data while increasing the temperature (denoted as ZFW-I) as shown in
Fig.~\ref{m-steps}(a) and (b). We find that the remnant magnetization (0.33 $\mu_B$/f.u at
10 K) decreases with increasing temperature. Correspondingly the resistivity
increases with increasing the temperature up to 65 K and decreases thereafter.
This temperature is very close to the $T_{FM}$ as shown in inset of Fig.~\ref{m-temp}(a).
These result suggest that the phase coexistence developed in the system at 10 K, due to
the application and removal of 70 kOe field, persists up to $\sim$65 K.

\subsection*{Observation of multi-level resistance states}

In the 2nd protocol (ZFW-II) rather than warming the sample continuously from 10 K to
200 K we heat the sample up to 35 K and perform the temperature cycling between 35 and 10 K
as shown in Fig.~\ref{m-steps}(c) (indicated by step-1). After the temperature cycle
we further increase the temperature from 35 K up to 45 K and repeat the temperature cycling
(indicated by step-2). Similarly we have performed temperature cycling at 52 K and
58 K indicated by step-3 and step-4, respectively. During the temperature cycling,
from 35 K, 45 K, 52 K, and 58 K to 10 K, the magnetization remains unchanged and mimics
the memory effect. We also performed the resistivity measurements using the same protocol
[see Fig. ~\ref{m-steps}(d)] and there is a one-to-one correspondence between
the magnetization and the resistivity data. Experimental data for sweep rate dependence,
effect of repeated cycling, time dependence, reproducibility of multilevel states
(see supplementary Figs.6--10) shows that the generated four resistance
states are robust by its design. Not only four but large number of resistance states can
be designed using the same protocol.

\subsection*{Physical origin of memory states}

\begin{figure*}
\includegraphics[width=0.75\textwidth]{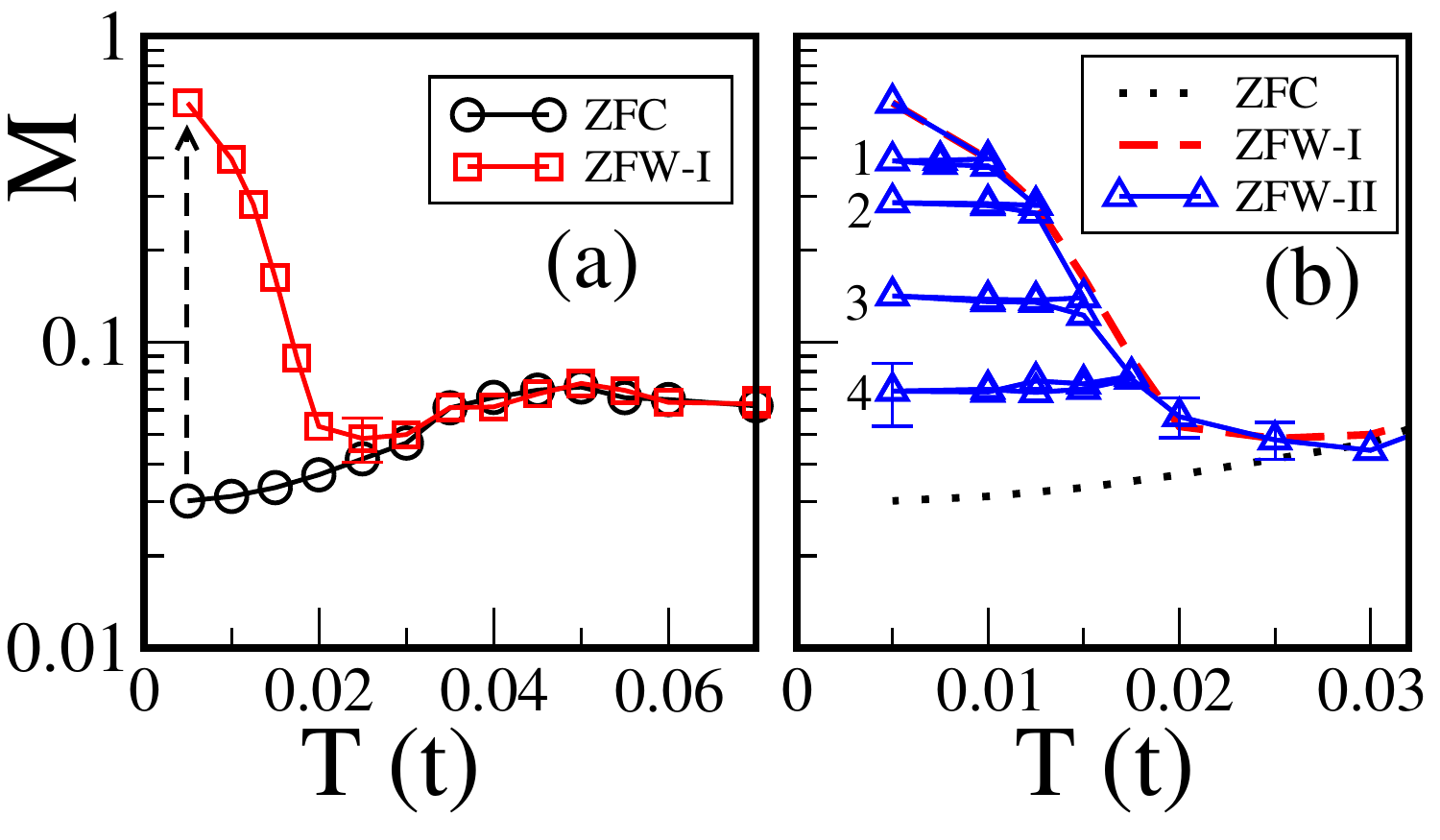}
\centering \caption[ ]
{Monte Carlo results: (a) Magnetization vs. temperature for ZFC and ZFW-I (with $h$=0)
protocols. Dashed arrow (connecting ZFC and ZFW-I) indicates application and removal
of magnetic field $h$=0.2 at T=0.005. Magnetization data calculated using ZFW-I
(using $h$=0) protocol is also plotted in (b) for comparison.
(b) Temperature dependence of magnetization using ZFW-II protocol. Temperature
cycling are performed at four representative temperatures [T=0.01 (step-1),
0.0125 (step-2), 0.015 (step-3), 0.0175 (step-4)].
Error bars are given wherever necessary. Error bars at all temperatures in step-4
are more or less same. So, error bar is given only at T=0.005
in step-4 for brevity.
}
\label{th_1}
\end{figure*}

\begin{figure*}
\includegraphics[width=0.95\textwidth]{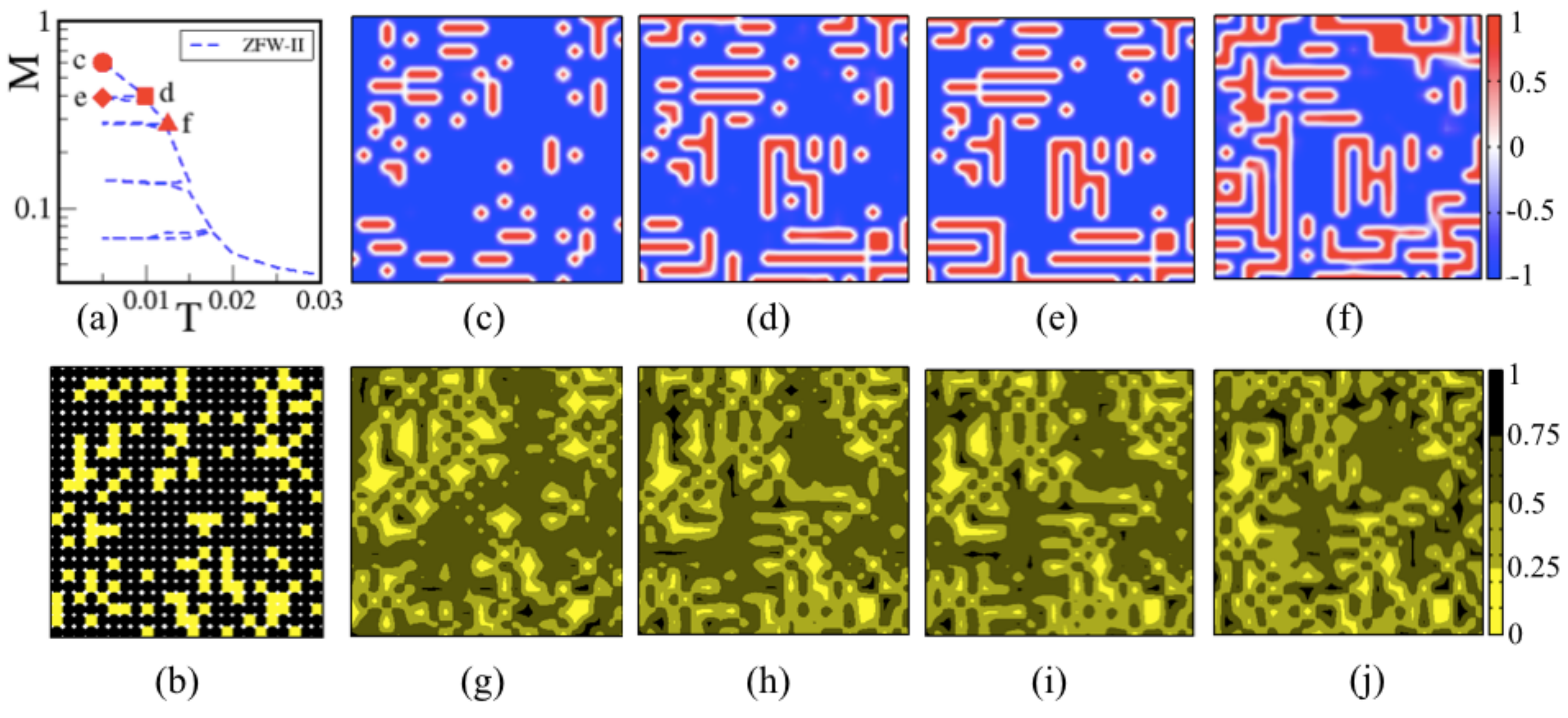}
\centering \caption[ ]
{(a) ZFW-II data from fig.~\ref{th_1}(b) is reproduced here highlighting four
temperature points using four different symbols; (b) disorder distribution [$\epsilon_i$
distribution] for one copy of disorder. Black (yellow) points represent
$\epsilon_i$ = -0.3 ($\epsilon_i$ = 0.3). (c)--(f): The $z$ components of simulated
Mn (t$_{\rm 2g}$) spins at each sites and (g)--(j): corresponding electron density at
each sites on a 24$\times$24 lattice using one of the disorder configuration
[shown in (b)]. (g) is the corresponding electron density for (c), (h) is the
corresponding electron density for (d) and so on. Color bar [right of (h)] is
for (c)--(f) and Color bar [right of (j)] is for (g)--(j).
}
\label{th_2}
\end{figure*}

\begin{figure*}
\includegraphics[width=0.75\textwidth]{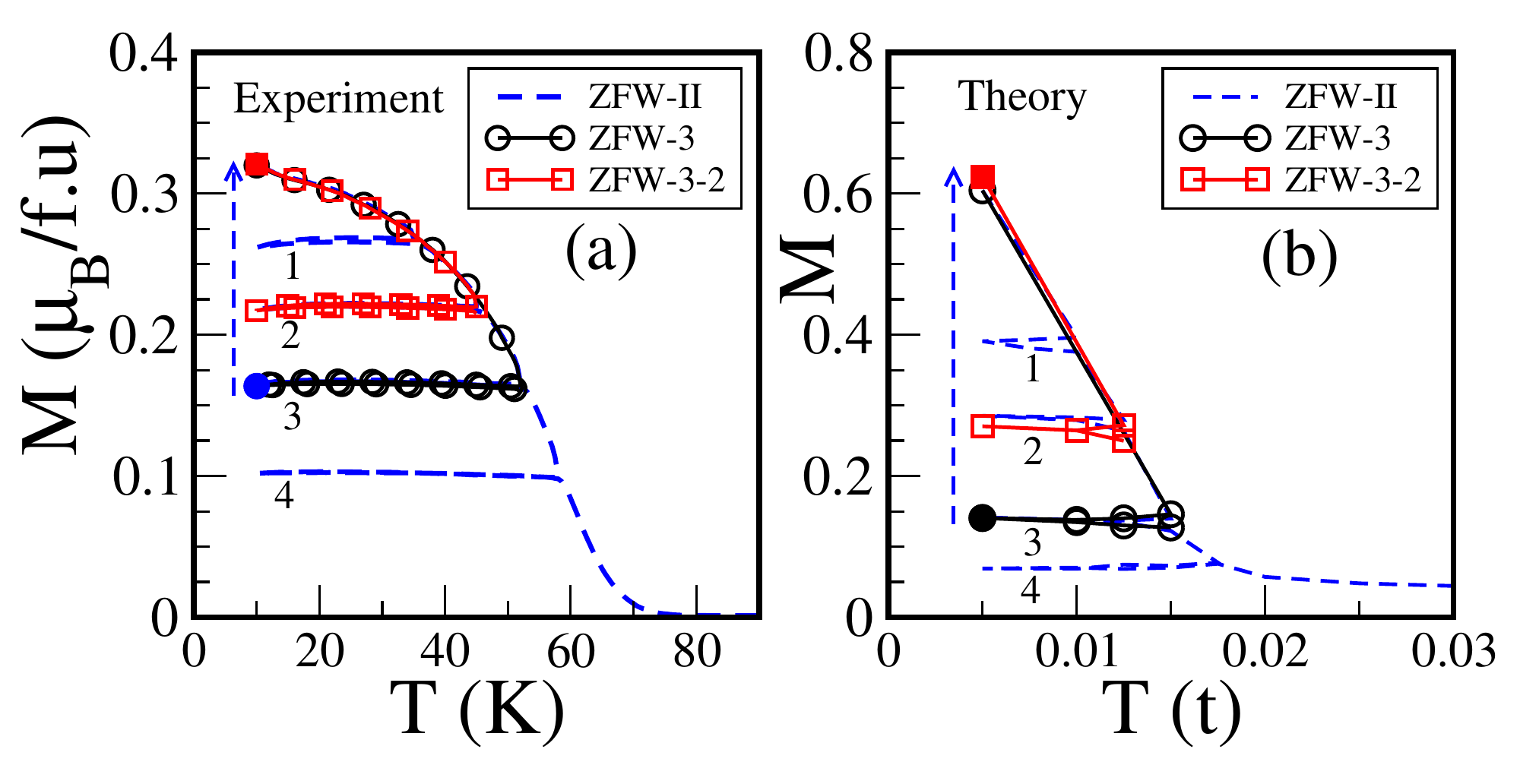}
\centering \caption[ ]
{Protocols to write, read, and erase multi-bit digital data: (a) Experiment: Dashed line is
using ZFW-II data reproduced from  Fig.~\ref{m-steps}(c). From 10 K we go to 52 K
(without any temperature cyclings at 35 K and 45 K) and perform temperature cycling
(i.e. going to step-3 directly, see ZFW-3). New step-3 data perfectly matches
with the step-3 data of ZFW-II. At T=10 K of new step-3 we apply and remove 70 kOe magnetic
field (i.e. going from filled circle symbol to filled square symbol, indicated by an arrow)
and then increase the temperature to 45 K. After that we perform a temperature cycling
below 45 K (see ZFW-3-2). The magnetization data from new step-2 also matches well with
the step-2 data of ZFW-II.
(b) Theory: Dashed line is using ZFW-II data reproduced from  Fig.~\ref{th_1}(b). Here
the magnetization data of step-3 is directly accessed without going through step-1 and
step-2. New step-3 data (see ZFW-3) perfectly matches with the step-3 data of ZFW-II.
Then $h$=0.2 is applied and removed at T = 0.005 of ZFW-3 (i.e. going from filled
circle symbol to filled square symbol, indicated by an arrow). After that we increase
the temperature to 0.0125 and perform a temperature cycling below it. The magnetization of
new step-2 data (see ZFW-3-2) also matches well with the step-2 data of ZFW-II.
Please see the text for more details.
}
\label{m-app}
\end{figure*}

During the temperature cycling, say at 35 K [see Fig. ~\ref{m-steps}(c)], it appears
that the size of the ferromagnetic regions freezes in a metastable state and as a
result the magnetization remains the same. In order to verify the presence 
of the metastable states we study a standard two-band model Hamiltonian for the manganites
in 2D. Due to the octahedral crystal field splitting the Mn $t_{\rm 2g}$ levels have
lower energy than the $e_{\rm g}$ levels and the itinerant $e_{\rm g}$ electrons are
coupled with the $t_{\rm 2g}$ electrons [$t_{\rm 2g}$ form the core Mn spins S (= 3/2)]
via large Hund's coupling. The $e_{\rm g}$ electrons are also coupled to the Jahn-Teller
phonons with a coupling strength $\lambda$. In addition, antiferromagnetic superexchange
interaction $J$ between neighboring core Mn spins is taken in to account. This well
investigated Hamiltonian in the double exchange limit~\cite{dagotto} reproduces the phase
diagram of manganites~\cite{yunoki,pradhanepl,pradhanprb}.
In SCSMO, ${Sr^{2+}}$ ions (larger in size as compared to $Sm^{3+}$ and $Ca^{2+}$ ions)
occupy the A-sites randomly and creates chemical disorder.
In order to incorporate this disorder $\sum_i \epsilon_i n_i$ is added at each Mn
site picked from the distribution
$P(\epsilon_i) = {1 \over 4} \delta(\epsilon_i - \Delta) + {3 \over 4} \delta(\epsilon_i + \Delta)$,
where $\Delta$ is the disorder potential.
In an external magnetic field $h$, a Zeeman coupling term $-\sum_i {\bf h}\cdot{\bf S}_i$
is added to the Hamiltonian, where ${\bf {S_i}}$ denote classical Mn t$_{\rm 2g}$ spin.
We use spin-fermion Monte Carlo (MC) technique based on the travelling cluster
approximation~\cite{tca-ref,pradhanprl} (TCA) for 24$\times$24 lattice
(see Ref.[\citenum{Sanjibnpg}] for details). We ensured that the system is well annealed
(using $10^4$ Monte Carlo system sweeps in general) at each temperature.

We measure $\lambda$, $J$, $\Delta$, $h$ and $T$ (temperature) in units of kinetic
hopping parameter $t$. The estimated value of $t$ in manganites is 0.2 eV~\cite{dagotto}.
We use $J$ = 0.1, $\lambda$ = 1.65 and $\Delta$=0.3 for SCSMO~\cite{Sanjibnpg}.
Magnetization [= $(S(\textbf{q}))^{0.5}$ at wave vector ${\bf q} = (0, 0)$,
where $S(\textbf{q})$ is magnetic structure factor
= ${1 \over 24^2}$ $\sum_{ij}$ $\bf {\bf S}_i\cdot {\bf S}_j$ e$^{i\bf{q} \cdot ({\bf r}_i-{\bf r}_j)}$]
is averaged over ten different disorder configurations in addition to the thermal
averages obtained during the Monte Carlo simulations.

Initially we cooled the system (from $T$ = 0.1 to $T$ = 0.005) and the corresponding
temperature dependence of the magnetization (M) for electron density $n$ = 1-$x$ = 0.5
is shown in Fig.~\ref{th_1}(a) using circle symbols. At $T$ = 0.005, we apply and remove
the external magnetic field ($h$ = 0.2). The magnetization of the system drops to 0.6
from 1 upon removal of the field. The temperature dependence of magnetization after
removing the magnetic field at T = 0.005 (defined as ZFW-I) is shown by the square
symbol in Fig.~\ref{th_1}(a). In the second
protocol (ZFW-II), similar to our experiment, after removing the field at T = 0.005,
we increase the temperature only up to 0.01 and perform a temperature cycling below
it [i.e. system is cooled to T = 0.005 and then heated up to T = 0.01] as
represented by step-1 in Fig.~\ref{th_1}(b)]. The magnetization remains unchanged
during the temperature cycle. Then from T = 0.01 we further increase the temperature
to 0.0125 and perform a temperature cycling below it. Similarly system is
recycled to lower temperature from T = 0.015 and T = 0.0175. In all cases the
magnetization remains unchanged during the temperature cycling.
So, our Monte Carlo calculations systematically reproduce the experimental results.

We move now to analyze the phase coexistence of ferromagnetic-metallic and
charge-ordered insulating phases using Monte Carlo snap-shots. Fig.~\ref{th_2}(a)
shows magnetization data using ZFW-II protocol (shown in Fig.~\ref{th_1}(b))
indicating four representative points. Figs.~\ref{th_2}(c)--(f) and
Figs.~\ref{th_2}(g)--(j) show the z components of $t_{2g}$ spins and electron density
for each sites for a disorder configuration, respectively. Fig.~\ref{th_2}(b) shows
$\epsilon_i$ distribution for that disorder configuration. At
T = 0.005 (after applying and removing $h$ = 0.2) the system consists of
ferromagnetic metallic and charge-ordered insulating [see Fig.~\ref{th_2}(c) and (g)]
phases. The electron density is more or less homogeneous ($\sim$0.65) within
the ferromagnetic clusters and charge-ordered (short-range) elsewhere.
Comparing with the impurity locations [Fig.~\ref{th_2}(b)] it is apparently clear that
the ferromagnetic-metallic clusters are within the homogeneously distributed $\epsilon_i$
regions and are pinned to the disorder configuration. This is the reason
for which the magnetization remains finite even after removing the magnetic field.
The size of the ferromagnetic-metallic clusters decreases in size with temperature
as shown in Figs.~\ref{th_2}(d) and (h). Interestingly going from T = 0.005 to T = 0.01
[i.e. from Fig.~\ref{th_2}(c) to Fig.~\ref{th_2}(d)] the magnetic regions get depleted
but remnant ferromagnetic regions at T = 0.01 are always part of the bigger ferromagnetic
regions seen in T = 0.005. During the temperature cycling from T = 0.01 to T = 0.005
(i.e. step-1) the ferromagnetic regions [compare Figs.~\ref{th_2}(d) and (e)] get
pinned in a metastable state due to the disorder configuration [shown in
Fig.~\ref{th_2}(b)] and as a result magnetization remains the same. For
T = 0.0125 the ferromagnetic regions (see Fig.~\ref{th_2}(f) and (j)) further
shrink in size and the magnetization decreases further.

\subsection*{Design of two-bit memory device}

We have purposefully designed four-level states to store a two-bit (00 or 01 or 10 or
11) digital data in each states. Say, we assign 00 to step-1, 01 to step-2, 10 to step-3
and 11 to step-4. Then the question we ask ourselves: can we write and erase the two-bit
data repeatedly in any of the four states? For this we go to 10 K [as discussed
in Fig.~\ref{m-steps}(a)] in the original sample. Then we apply and remove 70 kOe external
field, and increase the temperature to 52 K directly (i.e. going to step-3 directly) as
shown in Fig.~\ref{m-app}(a). Magnetization data from the thermal cycling at 52 K follows
the same path as shown in Fig.~\ref{m-steps}(c). This is to write two-bit information
(10) which can be read from resistivity measurement even after removing the temperature
probe. We have shown the one-to-one correspondence between magnetization resistivity in
Fig.~\ref{m-steps}. To erase the data we again apply and remove 70 kOe magnetic field at
10 K of step-3. Now in order to write another two-bit data, say 01, we increase the
temperature up to 45 K (i.e. step-2) and perform a thermal cycling. This step-by-step
procedure establishes a concrete way to write, read, and erase the two-bit digital data
in our four-level resistive system.

Similar to experiment we also increase the temperature to T = 0.015 (step-3) directly
from T = 0.005 (temperature at which magnetic field is applied and removed) by passing
two intermediate temperature cycles in our Monte Carlo calculations and perform a
temperature cycling below T = 0.015 [see Fig.~\ref{m-app}(b)]. The magnetization data
during temperature cycling at T = 0.015 perfectly matches with the step-3 of ZFW-II
(dashed line). During the temperature cycling from T = 0.015 we apply and removed
the magnetic field at T = 0.005 (i.e. going from filled circle symbol to filled
square symbol) to erase the data. Then to write a new data by using the step-2 we
increase the temperature directly to 0.0125 and perform a temperature cycling
below it. Magnetization data from new step-2 also matches well with the step-2
data of ZFW-II. So, our calculations reproduce the qualitative nature of the
experimental results.

We have used a direct temperature probe in our experiments. Temperature probe based
on the Joule heating~\cite{Lee}, by using voltage pulse, can be used to design the
device. The electronic phase separation (and in turn resistive states) can also be
controlled by numerous ways like, applying substrate strain, using mode-selective
vibrational excitations, using photo-induction and applying
pressure~\cite{Gillaspie,elovaara,Chai,Rini,Baldini} etc.

\subsection*{Conclusion}

In summary, our combined experimental and theoretical study outline a simple yet
effective path to achieve  multi-level resistance states that can give new direction
to architect multi-bit storage devices for future spintronic applications. Here we
devised the phase coexistence in a controllable manner in a half doped manganite below
60 K and showed that a temperature probe can be used to access four distinct metastable
phases for designing a two-bit memory device. In principle it can be extended to generate
many more distinctly resistive/magnetic states to devise more than two-bit devices. At
the end we demonstrate the writing, reading and erasing mechanism in our prototype
two-bit device. It is expected that our results will motivate experimenters to explore
electronic phase separated materials to engineer future multi-bit memory devices.

\subsection*{Methods }
\subsubsection*{Sample preparation}
High quality polycrystalline ${Sm_{0.5}(Ca_{0.5}Sr_{0.5})_{0.5}MnO_3}$ compound was
prepared by the well-known sol-gel technique by taking appropriate amount of $Sm_2O_3$,
$CaCO_3$, $Sr(NO_3)_2$ and $MnO_2$ as the starting materials of purity 99.9\%. To prepare
the bulk polycrystalline sample the decomposed gel was palletized and subsequently heated
for 36 h at 1300$^0$C. For more details (also for crystalline structure study) please see
Ref.~\citenum{Sanjibnpg}.

\subsubsection*{Magnetic property measurements}
Magnetic properties was measured by utilizing Superconducting Quantum Interference Device
Magnetometer (SQUID-VSM) of Quantum Design with maximum magnetic field value of 70 kOe
in the temperature range 10 K - 300 K.

\subsubsection*{Electrical and magneto-transport measurement}
The electrical transport and magnetotransport measurements were performed by four probe
method using the longitudinal geometry of the bar-shaped samples. Zero field and in field
resistance measurements in the temperature range 20 K to 300 K was carried out using
Keithley source and measure unit 2651A  with the measurement limit of $10^{11}$ ohm.
Below 20 K, the zero field resistance measurement was extended by using Keithley
Electrometer 6517A in the capacitor arrangement method and at 10 K it exceeds our
measurements limit ($10^{13}$ ohm).

\section*{References}

\section*{Acknowledgements}
The work was supported by Department of Atomic Energy (DAE), Govt. of India.
We thank Sudhakar Yarlagadda for his comments on the manuscript.

\section*{Author contributions}
ID and KP developed the concept of the study. SB and KD prepared the samples.
SB performed the most of the experimental work. KD and ID performed partial measurements
and KP performed the numerical work. SB and KP wrote the draft of the paper and all authors
reviewed the manuscript.

\pagebreak[4]
\section{Supplementary Information}                                         
\begin{figure}[h!]
\includegraphics[width=0.35\textwidth]{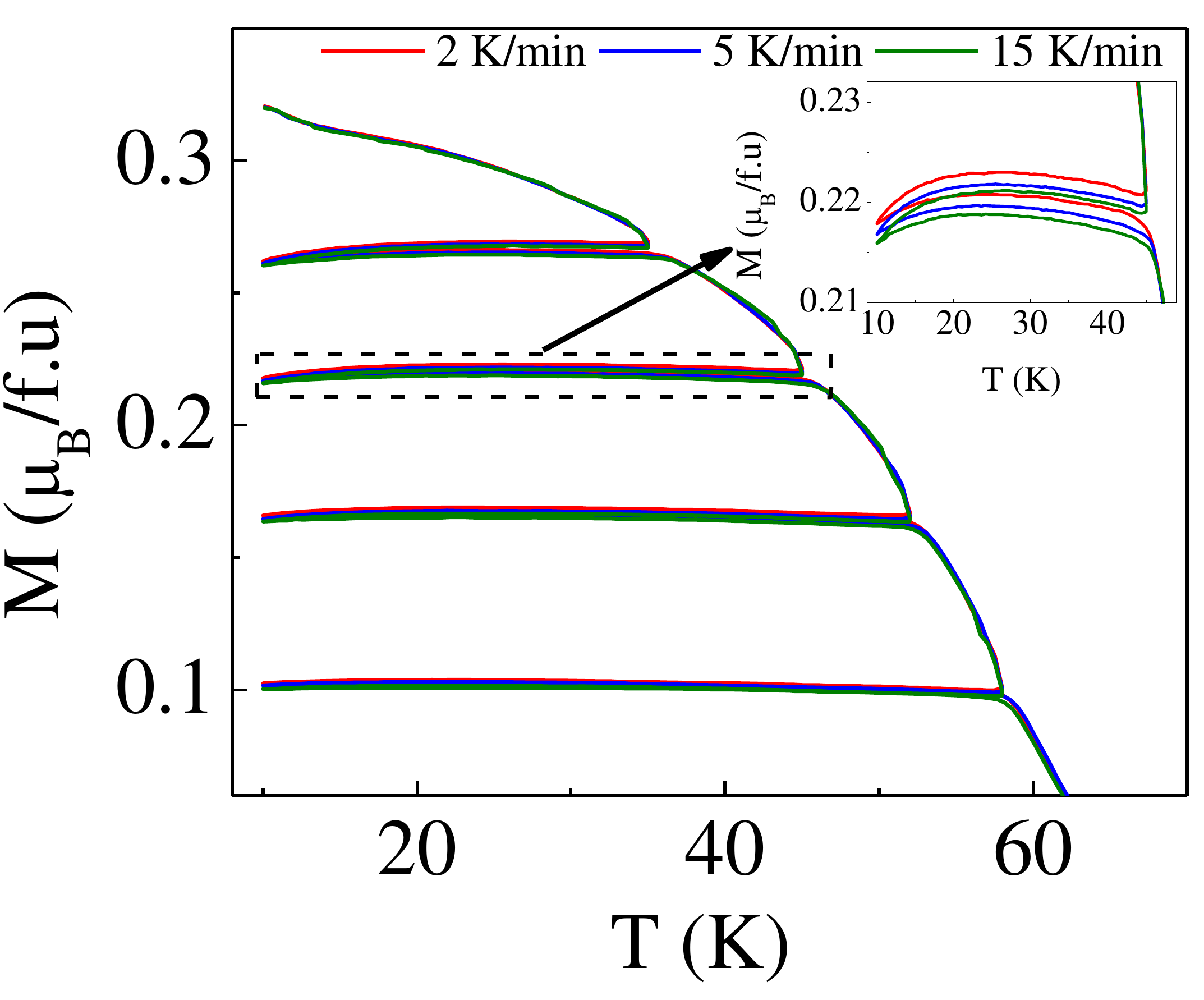}
\centering
\caption[]{Sweep rate dependence (supplementary figure):
Temperature dependence of magnetization data for the different sweep rates 2 K/min,
5 K/mim and 15 K/min for ZFW-II protocol (see the main text). Inset in the figure
shows the zoomed portion of one of the step.}
\end{figure}

\begin{figure*}
\includegraphics[width=0.7\textwidth]{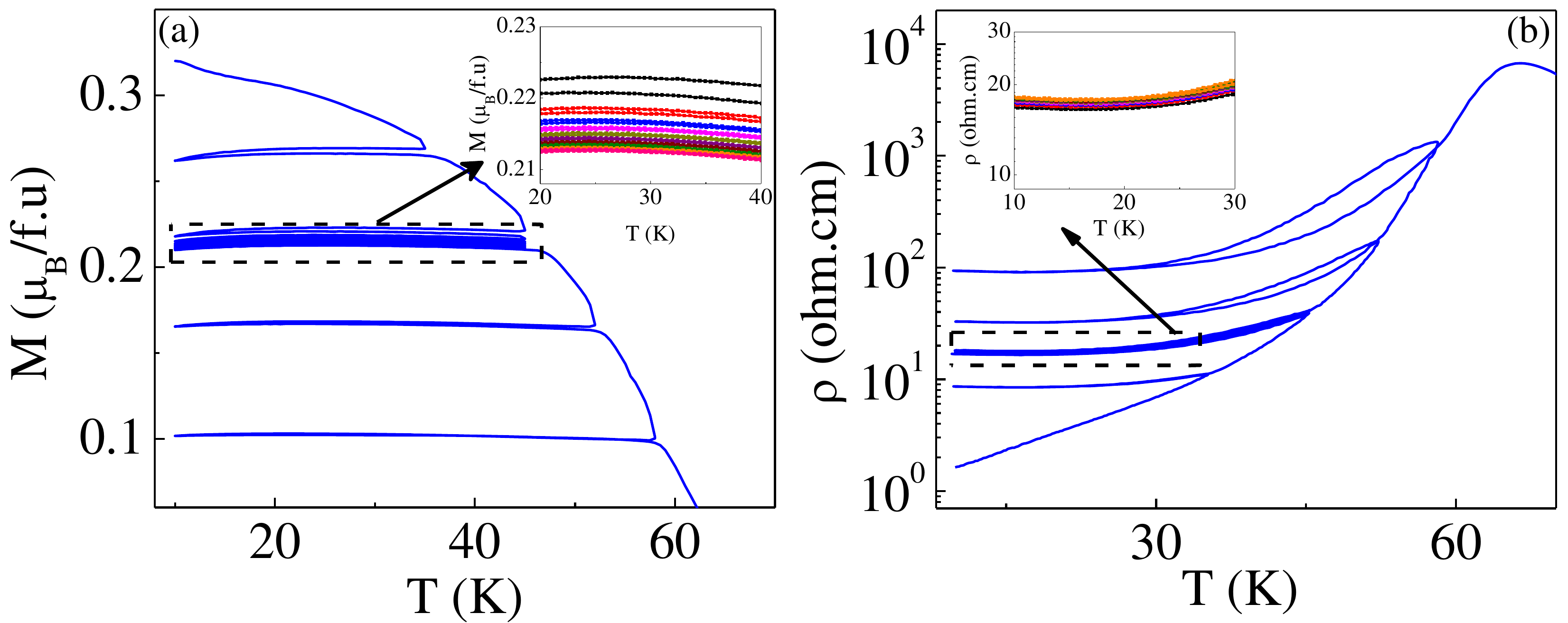}
\centering
\caption[]{Effect of repeated cycling (supplementary figure):
Temperature dependent (a) magnetization [(b) resistivity] data for 5 K/min using
protocol ZFW-II where 10 temperature cycling are performed in step-2. Insets show
the zoomed portion of step-2 data. Experimental data remains unaffected with repeated
cycling.
}
\end{figure*}

\begin{figure*}
\includegraphics[width=0.7\textwidth]{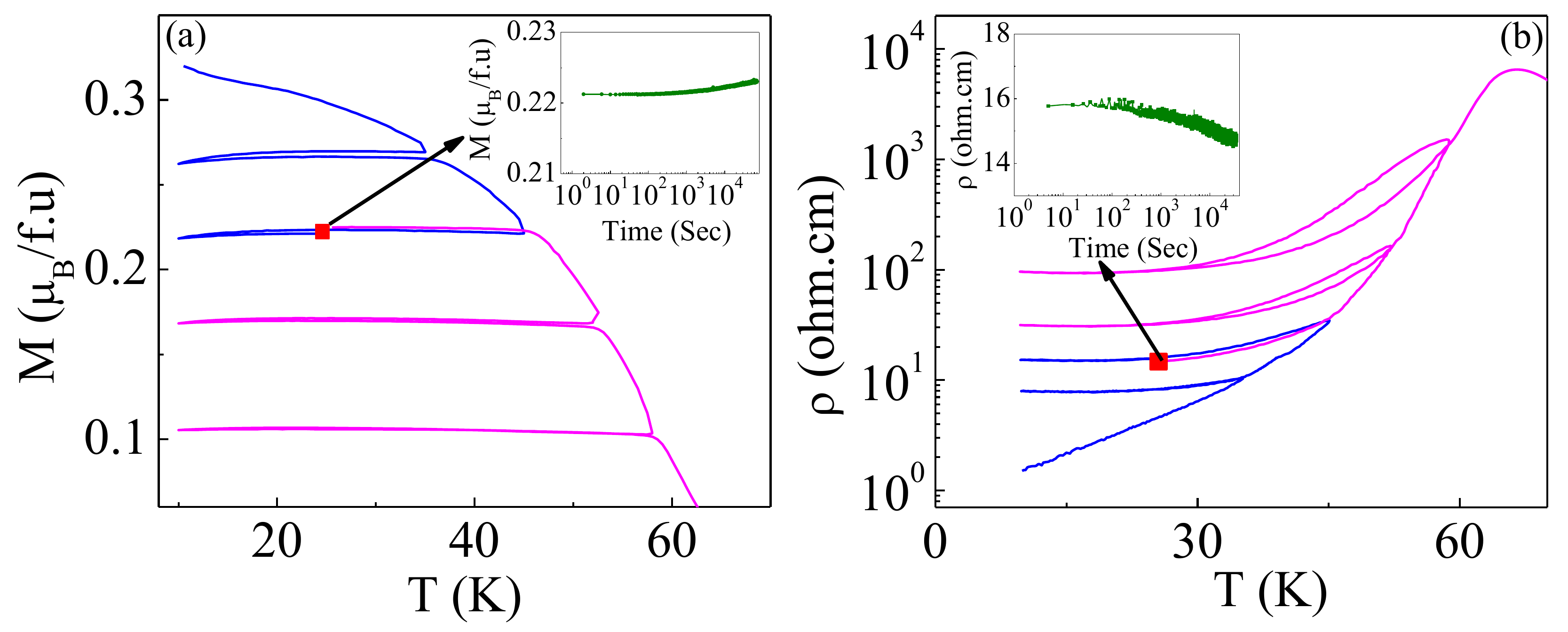}
\centering
\caption[]{Time dependence-I (supplementary figure): (a) M(T) and (b) $\rho$ (T)
measured in the ZFW-II protocol. Magnetization [in (a)] and resistivity [in (b)]
data are taken for 10 hours at 25 K of step-2 to see the time evolution [see the
inset (a) and (b)]. Magnetization and resistivity remains more or less same even
after 10 hours.
}
\end{figure*}

\begin{figure*}
\includegraphics[width=0.7\textwidth]{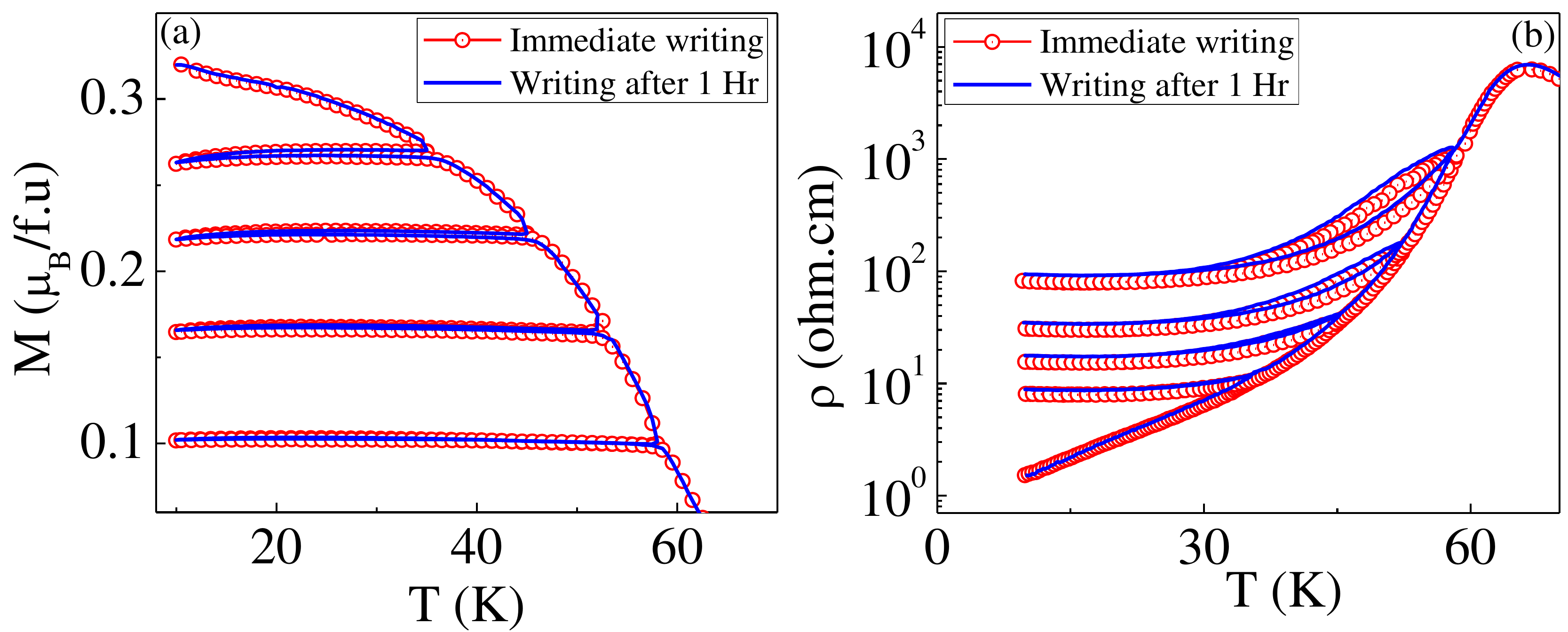}
\centering
\caption[]{Time dependence-II (supplementary figure):
(a) M(T) and (b) $\rho$ (T) measured using ZFW-II protocol.
In one case magnetization data is taken immediately after removing the magnetic
field at 10 K (see main text) while in other magnetization data is taken one hour
later after removing the field. Resistivity data are also taken in a similar fashion.
This shows that system do not loose its identity with time.
}
\end{figure*}

\begin{figure*}
\includegraphics[width=0.7\textwidth]{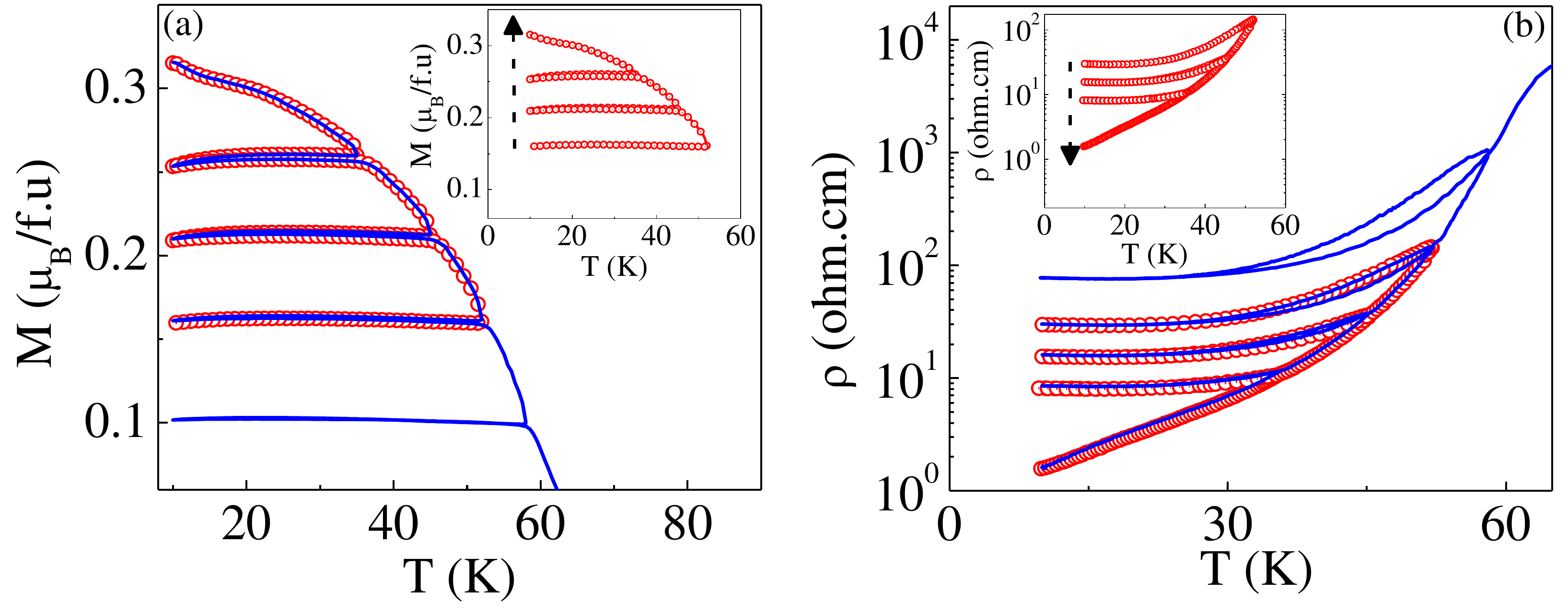}
\centering
\caption[]{Reproducibility of multilevel states (supplementary figure): Insets of (a)
and (b) show repetition of Fig. 4(a) and (b), but the we stopped at 10 K of step '3'.
Then we apply and remove a 70 kOe magnetic field (shown as dotted arrow). Then the
magnetization and resistivity with temperature [using ZFW-II protocol] are measured
and plotted in main Figs. (a) and (b),
repectively using blue solid lines. Line with red circles in the main plot are from
the inset. This show that all the four steps can be obtained repeatedly.}
\end{figure*}

\end{document}